# Education and Outreach in Particle Physics

R. Michael Barnett
*Lawrence Berkeley National Laboratory, Berkeley, CA 94720, USA*

There are many varied programs of education and outreach in particle physics. This report for the Division of Particles and Fields of the American Physical Society 2001 meeting reviews the impact of these programs in general, and also gives several examples of ongoing programs with a primary focus on those in the US.

## 1. Introduction

In this talk, I discussed how the outside world views us, and what we can communicate that students and the public will appreciate and retain. Our goal should be to inspire and engage students and the public.

## 2. A Sense of Wonder

Former President Bill Clinton, spoke at the World Economic Summit in Davos in January 2011. For most of the hour, the moderator asked him about the past and present situation in the world. He then asked about the next 20 years, and Clinton immediately turned to science, and in particular, the Large Hadron Collider. It is notable that science was the theme of his remarks about the future. While the details about the LHC were fuzzy, what he emphasized most was that we impart a “sense of wonder”. I believe that this is what we are most able to convey with our outreach efforts.

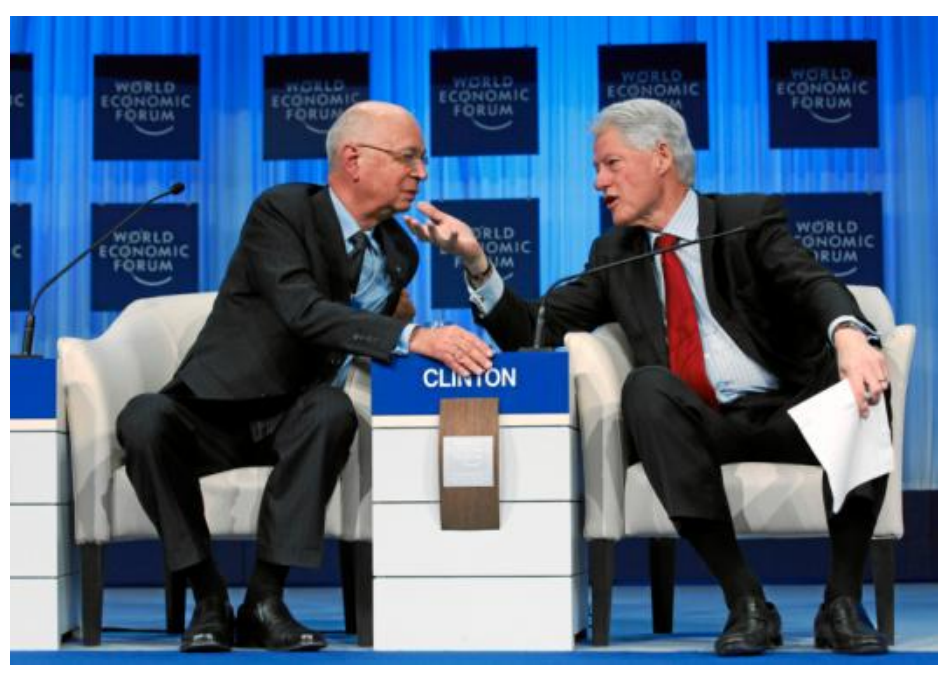

Figure 1: Former President Bill Clinton speaking in Davos 2011

A similar theme came from a new National Research Council study led by Helen Quinn. It proposed a new framework for improving American science education. The report said that one of the big goals was to “ensure that by the end of 12th grade, all students have some appreciation of the beauty and wonder of science”. The new framework calls for paring the curriculum to focus on core ideas and teaching students more about how to approach and solve problems rather than just memorizing factual nuggets.

## 3. How does Hollywood Present Particle Physics?

There are many examples in literature, art and film of depictions of particle physics and/or particle physicists. In addition, we have feedback from Hollywood about our field. A recent example came from director and star of the major Hollywood film Angels & Demons, Ron Howard and Tom Hanks, commenting on LHC. This can be seen in the extra feature on the DVD of Angels & Demons. In addition, there are comments in:

http://atlas.ch/angels/tom-hanks-clip.html and
http://atlas.ch/angels/ron-howard-clip.html

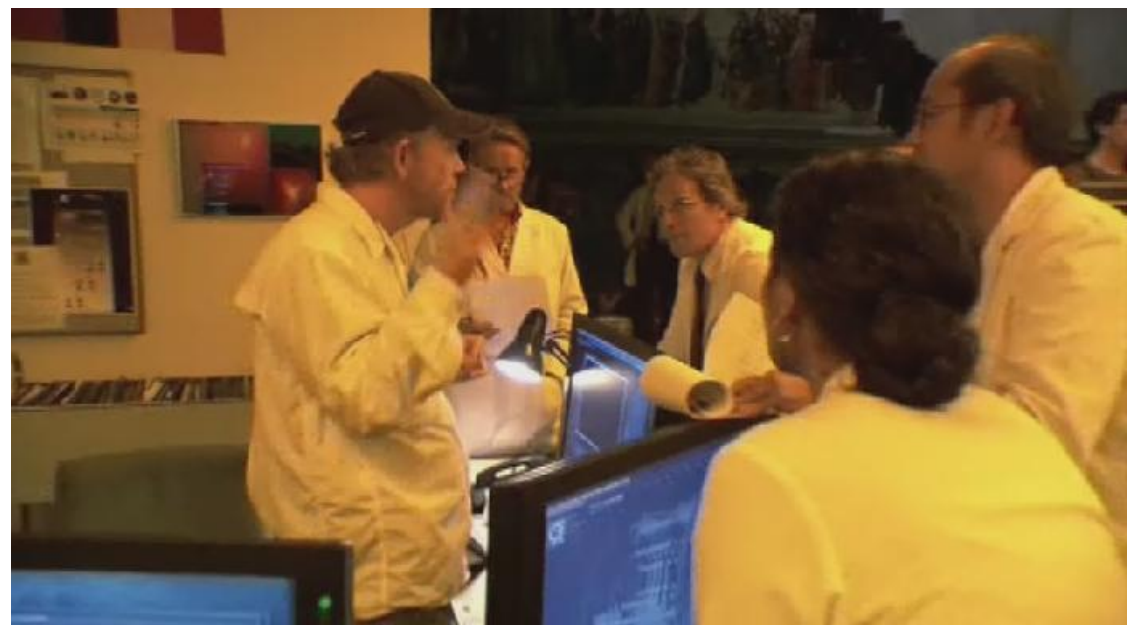

Figure 2: Ron Howard commenting on the LHC

When Leon Lederman introduced the term “God Particle” for the Higgs boson in his book by that title, he could not have foreseen the impact in the media of this expression.

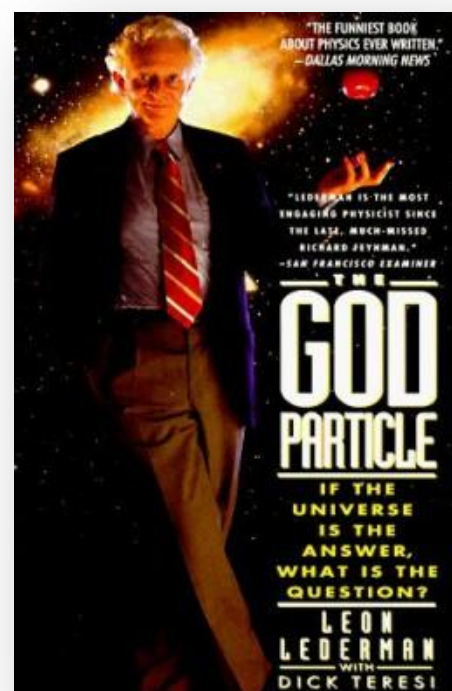

Figure 3: Cover of Leon Lederman’s book

Not only has this expression appeared in innumerable news stories in newspapers and television, but also in movies such as Angels & Demons, where both antimatter and the God particle are discussed, see: [http://www.atlas.ch/angels/god-particle.html](http://www.atlas.ch/angels/god-particle.html)

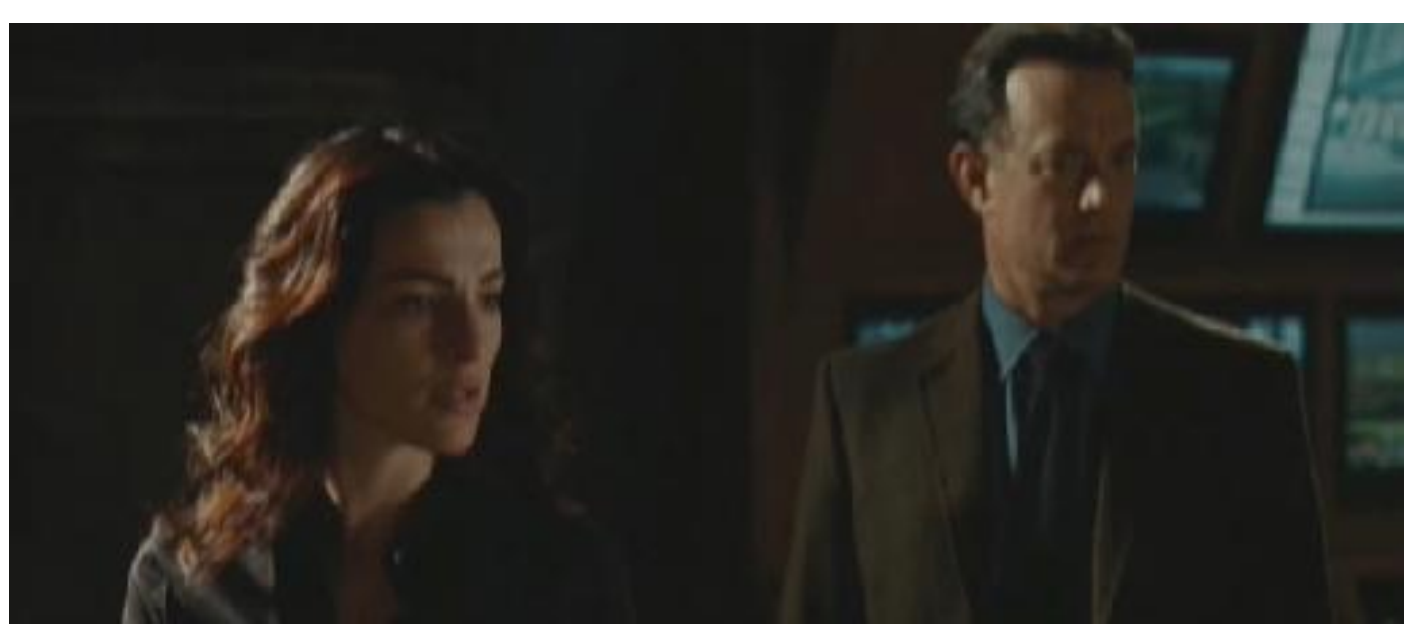

Figure 4: Actress Ayelet Zurer explaining the God particle in Angels & Demons
(TM and © Columbia Pictures Industries, Inc.)

## 4. Three Views of Accelerators and Collisions

I think it is informative to look at three contrasting views of accelerators and collision events. The first is a recent example from the New York Times.

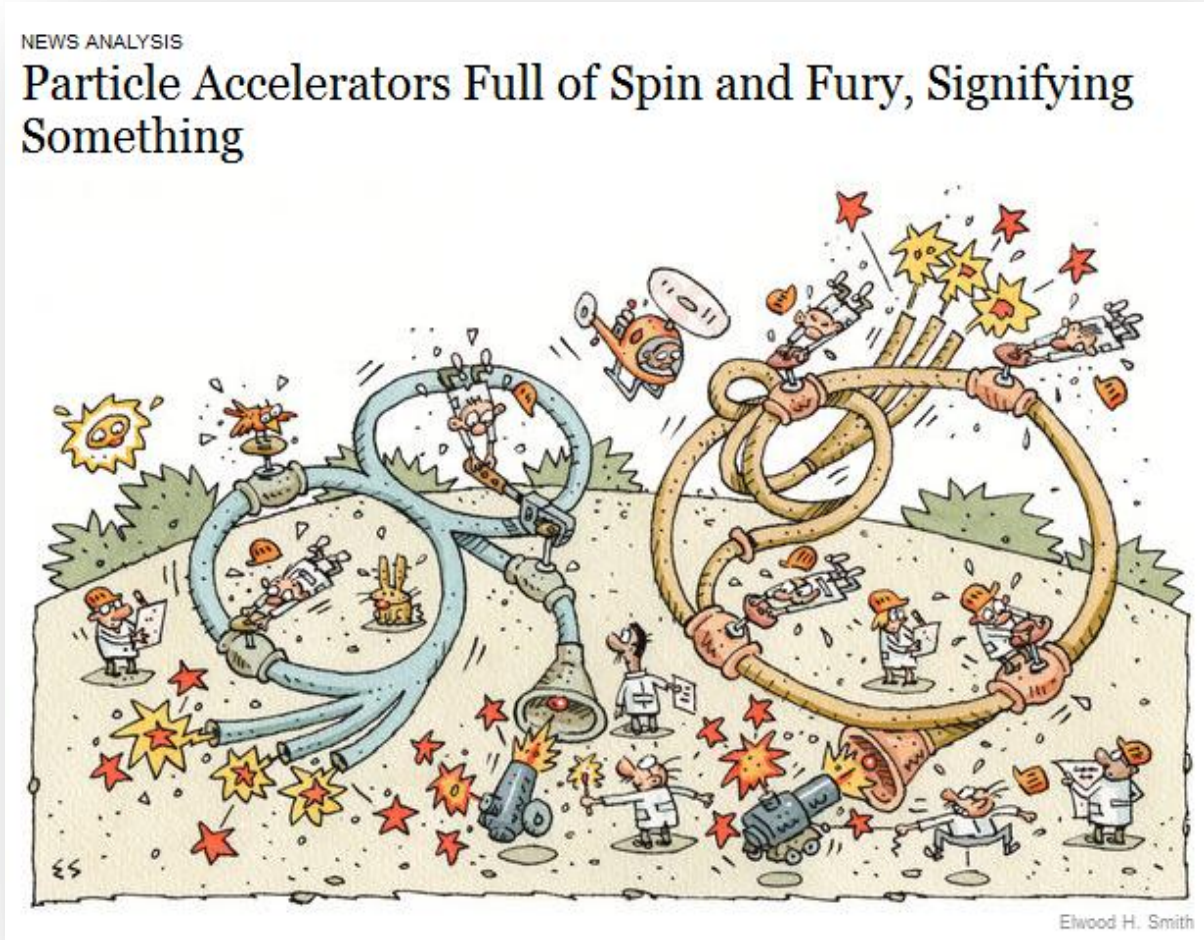

Figure 5: An illustration of accelerators from the New York Times

The Outreach program of the ATLAS Experiment produced an animation of the LHC and then a collision within ATLAS, which can be seen at: http://www.atlas.ch/multimedia/actual-events.html

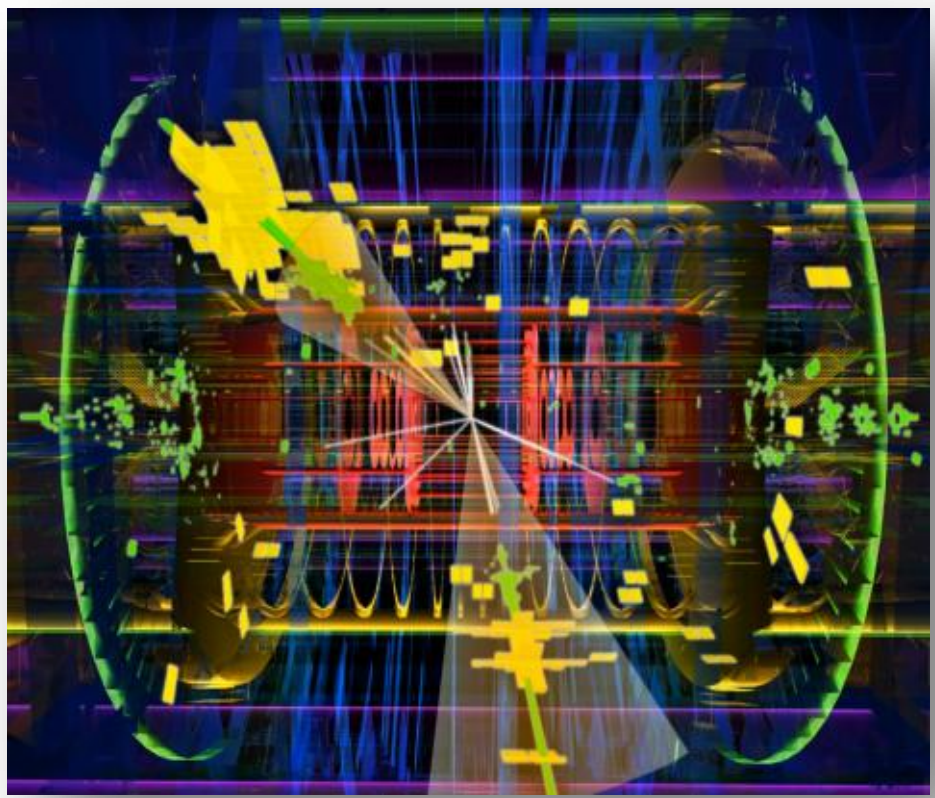

Figure 6: A depiction of acceleration and collision produced by the ATLAS Experiment

The film Angels & Demons also produced a similar depiction, which can be seen in the DVD.

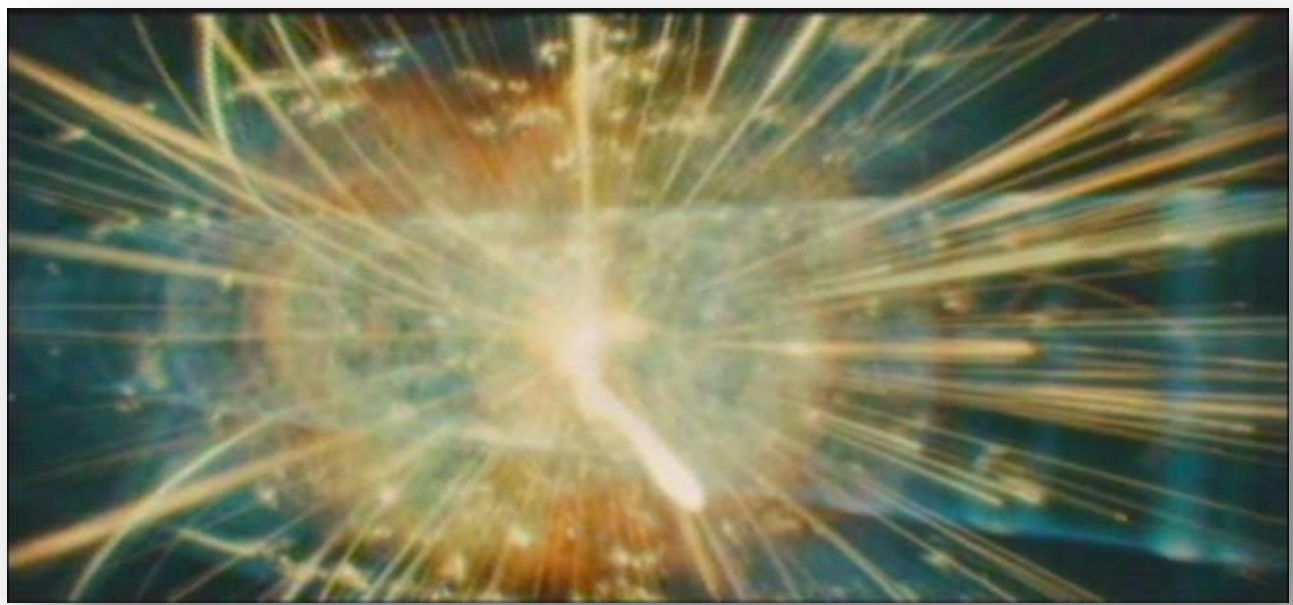

Figure 7: A depiction of acceleration and collision in Angels & Demons
(TM and © Columbia Pictures Industries, Inc.)

## 5. Science in Comics (and Animation of Comics)

A very nice example of the use of comics to communicate science appears in the comic called PhD Comics. Some of these are also animated, see an example about dark matter at: http://www.phdcomics.com/comics.php?f=1430

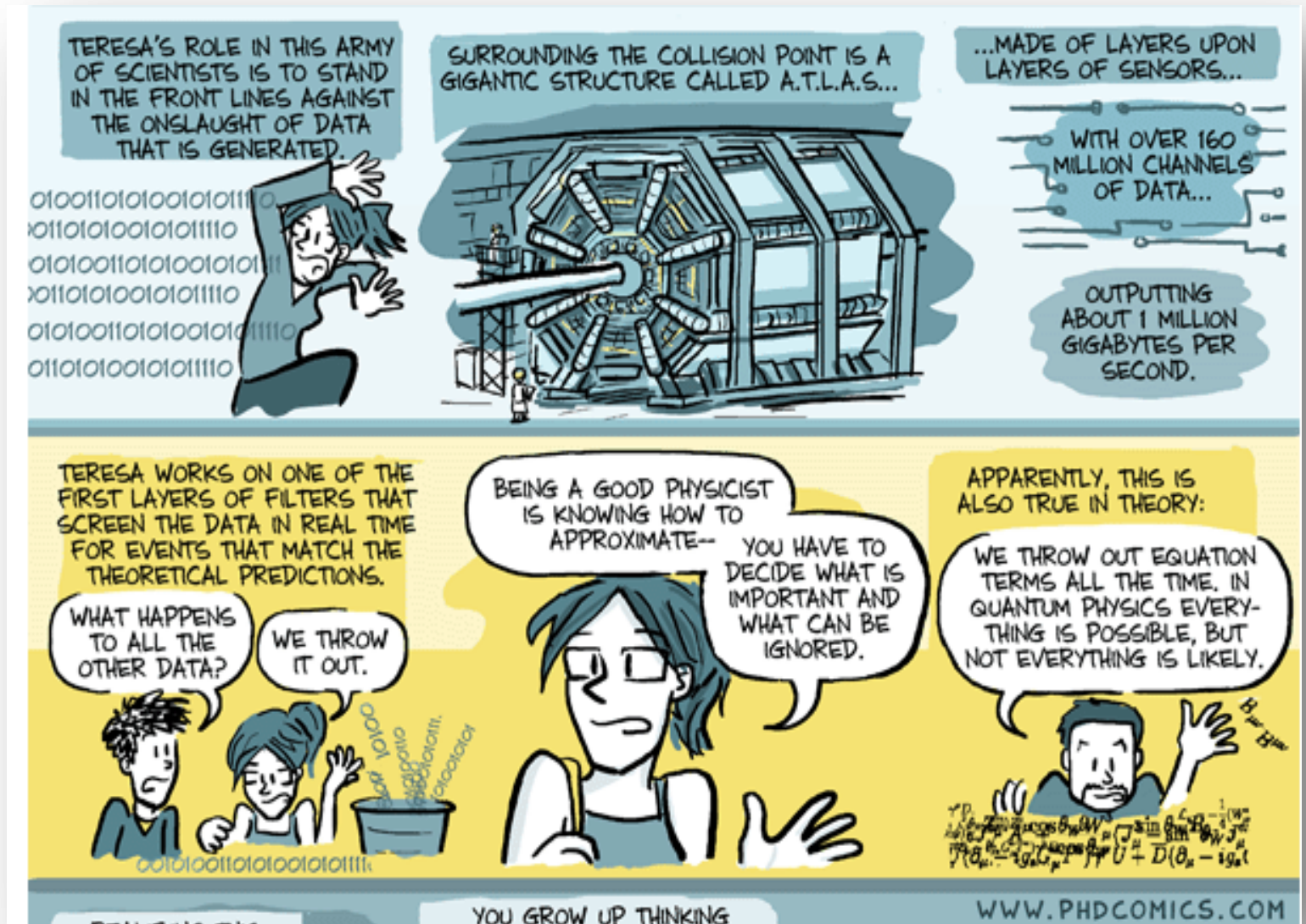

Figure 8: Example of a comic from PhD comics by Jorge Cham

## 6. Unexpected Impacts

The impacts of our science can be unexpected. I have collected just a few examples. The opera Les Troyens by Berlioz, shown in Valencia, St. Peterburg and Warsaw, used a set design based on the ATLAS detector.

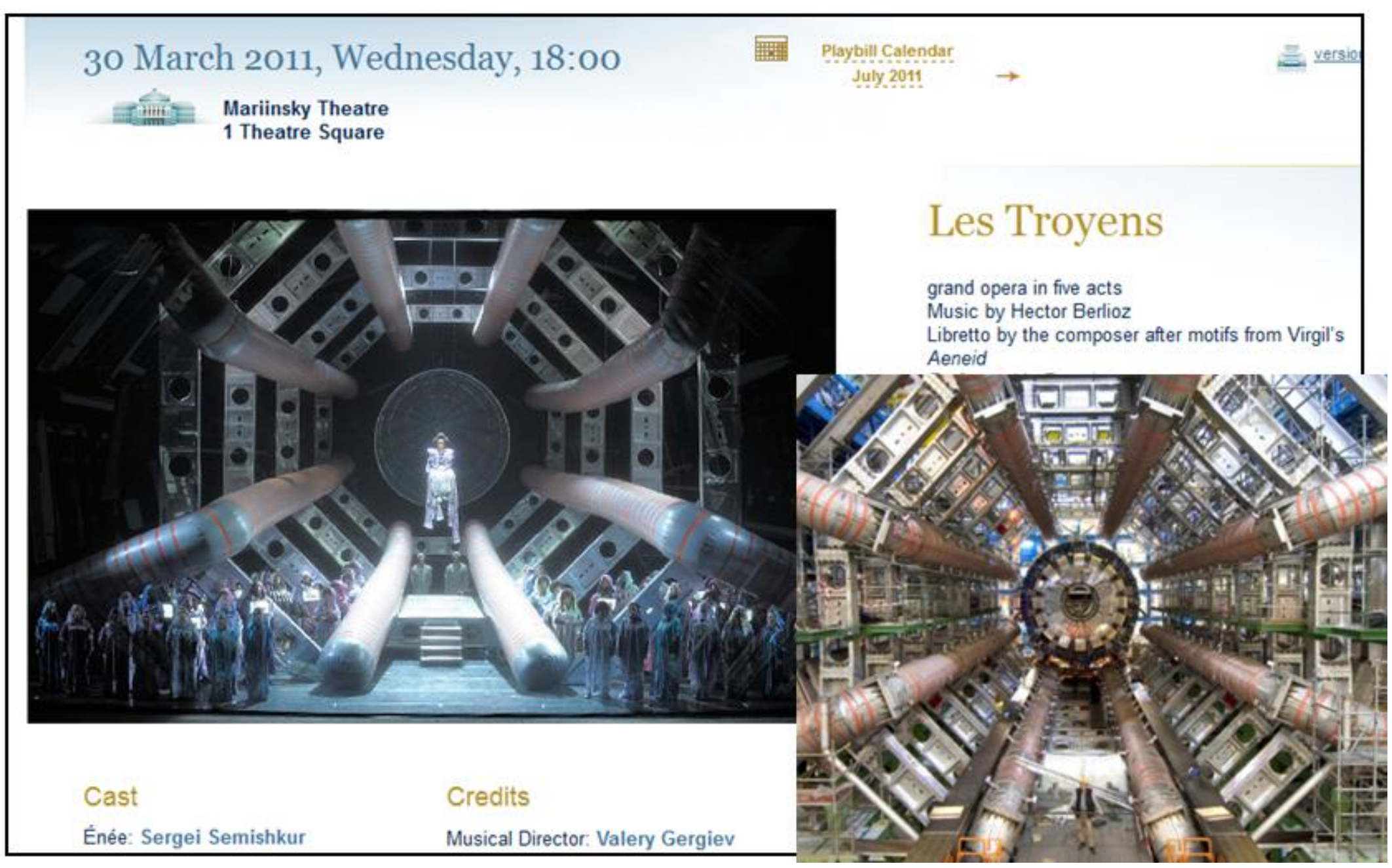

Figure 9: A set design for Les Troyens based on the ATLAS detector (inset photo).

The forthcoming Muppets movie (due out in November 2011) apparently has a scene appearing to be taken in the ATLAS cavern. This we only know from the trailer: http://www.youtube.com/watch?v=C4YhbpuGdwQ

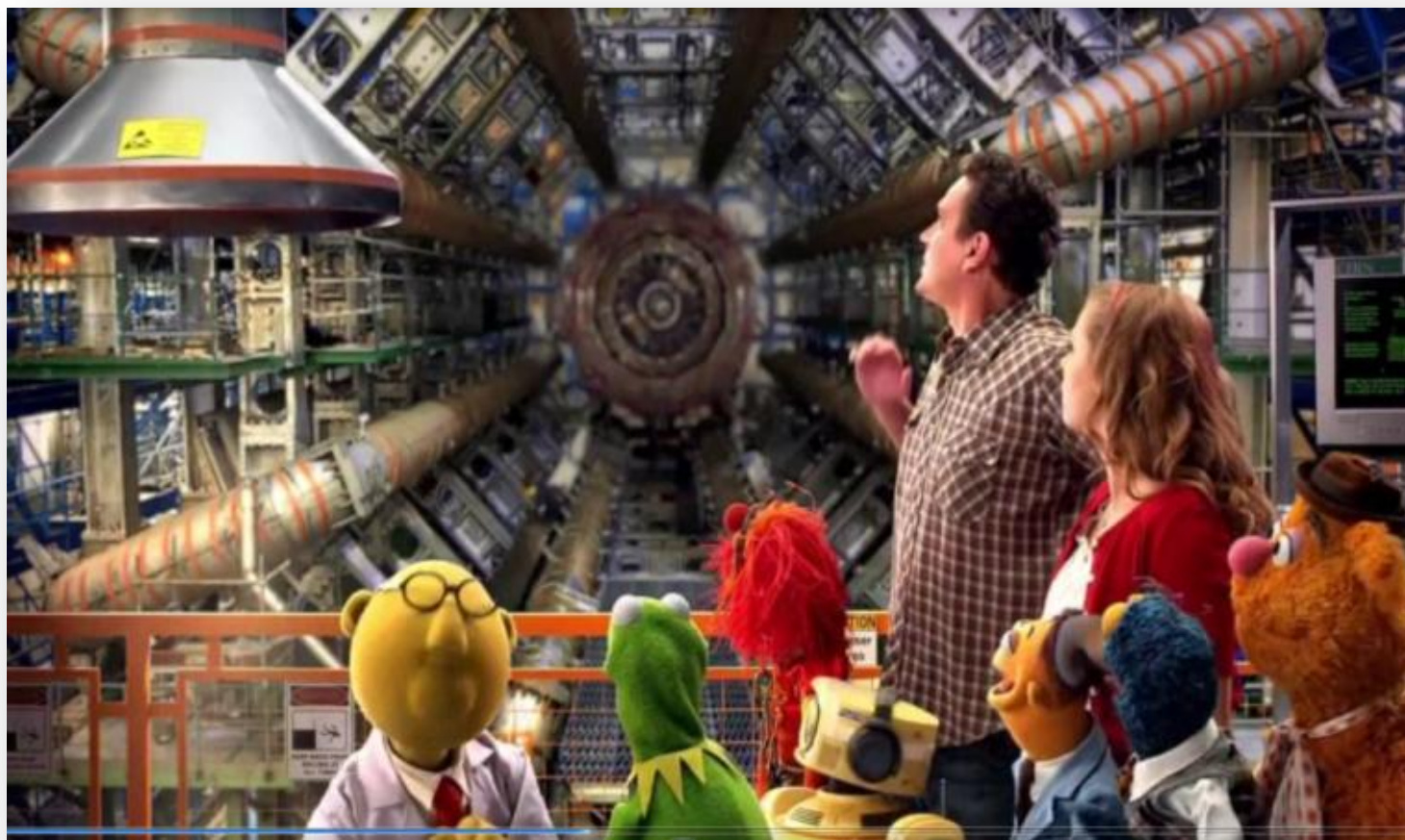

Figure 10: From the Muppets movie trailer on YouTube

For several years the charts produced by the Contemporary Physics Education Project (http:// CPEPphysics.org) have been appearing in movies and television. The movies include: Donnie Darko, Thirteen Conversations about One Thing, and Hulk. On television, the charts can be seen on Big Bang Theory.

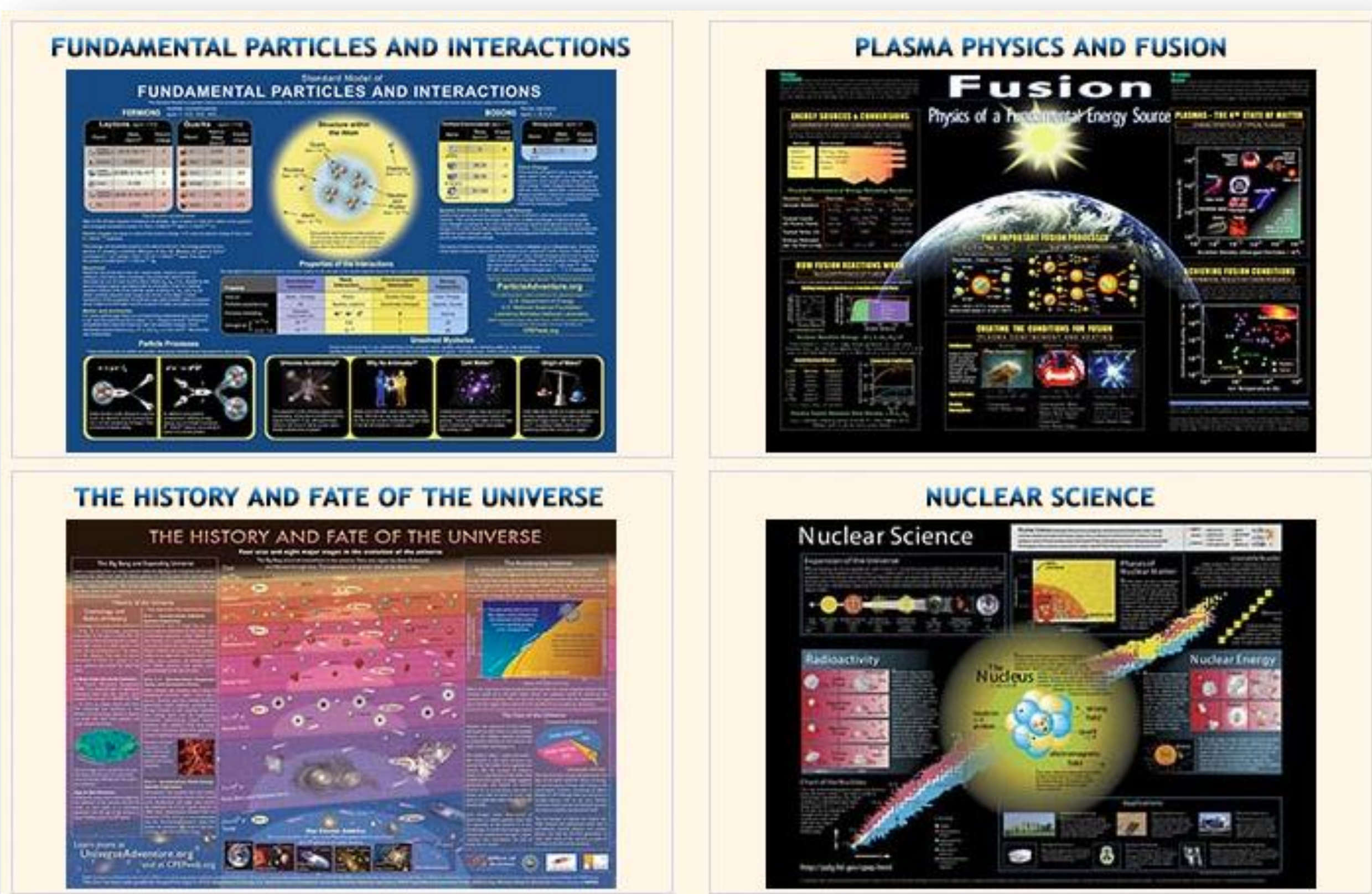

Figure 11: The charts of the Contemporary Physics Education Project

Popular Science magazine recently had an article on the “Universe’s Ten Most Epic Projects” and the Large Hadron Collider was number two.

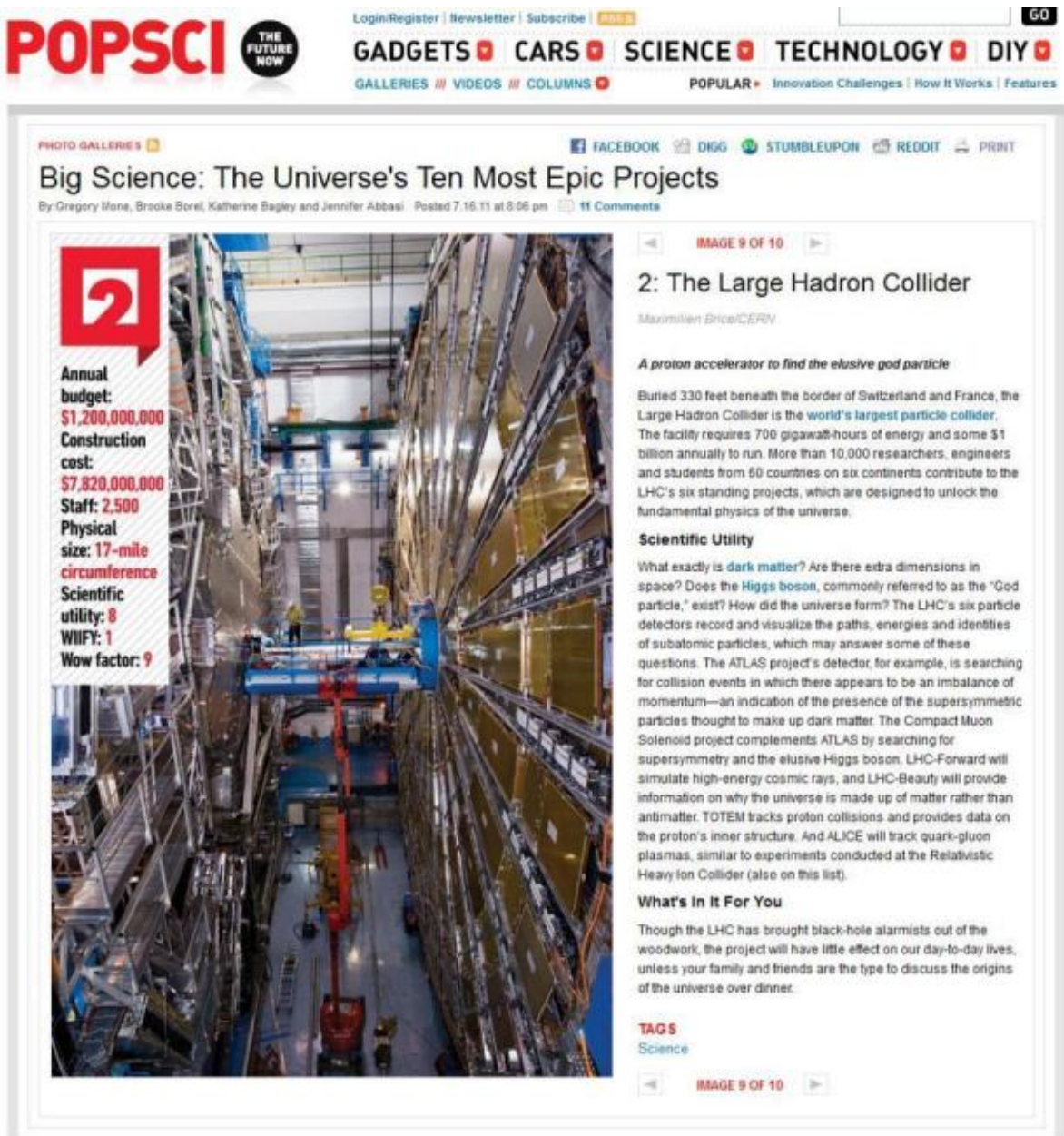

POPSCI THE FUTURE NOW

Login/Register | Newsletter | Subscribe

GADGETS | CARS | SCIENCE | TECHNOLOGY | DIY

GALLERIES /// VIDEOS /// COLUMNS

POPULAR ▸ Innovation Challenges | How It Works | Features

PHOTO GALLERIES

FACEBOOK DIGG STUMBLEUPON REDDIT PRINT

Big Science: The Universe's Ten Most Epic Projects

By Gregory Mone, Brooke Borel, Katherine Bagley and Jennifer Abbasi Posted 7.16.11 at 8:06 pm 11 Comments

2

Annual budget: $1,200,000,000
Construction cost: $7,820,000,000
Staff: 2,500
Physical size: 17-mile circumference
Scientific utility: 8
WIIFY: 1
Wow factor: 9

IMAGE 9 OF 10

2: The Large Hadron Collider

Maximilien Brice/CERN

*A proton accelerator to find the elusive god particle*

Buried 330 feet beneath the border of Switzerland and France, the Large Hadron Collider is the world's largest particle collider. The facility requires 700 gigawatt-hours of energy and some $1 billion annually to run. More than 10,000 researchers, engineers and students from 60 countries on six continents contribute to the LHC's six standing projects, which are designed to unlock the fundamental physics of the universe.

**Scientific Utility**

What exactly is dark matter? Are there extra dimensions in space? Does the Higgs boson, commonly referred to as the "God particle," exist? How did the universe form? The LHC's six particle detectors record and visualize the paths, energies and identities of subatomic particles, which may answer some of these questions. The ATLAS project's detector, for example, is searching for collision events in which there appears to be an imbalance of momentum—an indication of the presence of the supersymmetric particles thought to make up dark matter. The Compact Muon Solenoid project complements ATLAS by searching for supersymmetry and the elusive Higgs boson. LHC-Forward will simulate high-energy cosmic rays, and LHC-Beauty will provide information on why the universe is made up of matter rather than antimatter. TOTEM tracks proton collisions and provides data on the proton's inner structure. And ALICE will track quark-gluon plasmas, similar to experiments conducted at the Relativistic Heavy Ion Collider (also on this list).

**What's In It For You**

Though the LHC has brought black-hole alarmists out of the woodwork, the project will have little effect on our day-to-day lives, unless your family and friends are the type to discuss the origins of the universe over dinner.

TAGS
Science

IMAGE 9 OF 10

Figure 12: Article in Popular Science magazine

I noted that following the EPS-2011 meeting, the 9th most popular article on BBC was the Higgs boson news, following then headline news about attacks on students in Norway, debt talks in Washington, and coverage of the phone hacking scandal in Britain. More recently the neutrino speed story was the lead story on the Huffington Post website (on 23 September 2011).

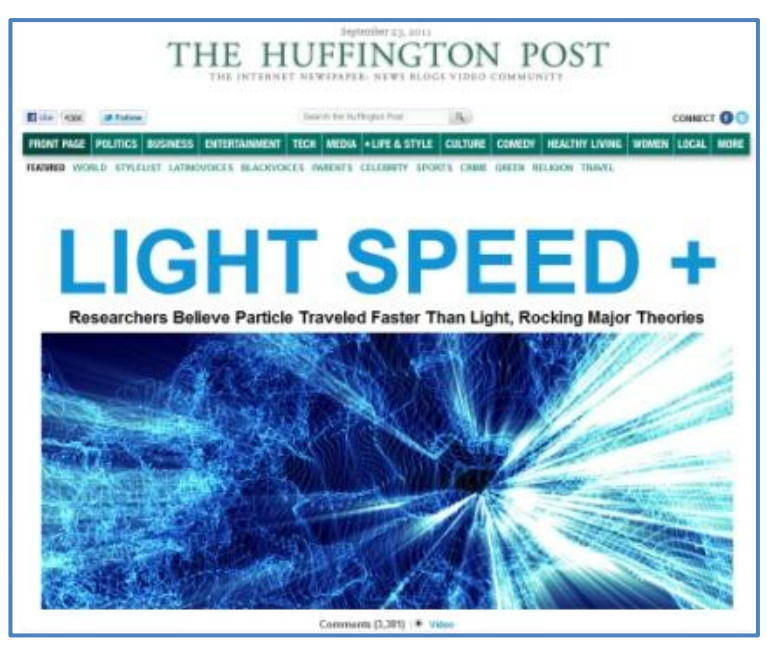

Figure 13: Huffington Post home page from 23 September 2011

## 7. Education and Outreach Programs across HEP

This section is based on the talks (and slides) of the parallel session speakers. I am not reporting on the extensive programs by ATLAS and CMS (see their websites for details): http://atlas.ch and http://cms.cern.ch

## 7.1. Education at Fermilab

Marge Bardeen of Fermilab reported on the extensive Education programs at Fermilab. These programs are a legacy of Leon Lederman. Leon began Saturday Morning Physics in 1980 to "use the magnificence of Fermilab to dazzle (and capture) high school kids."

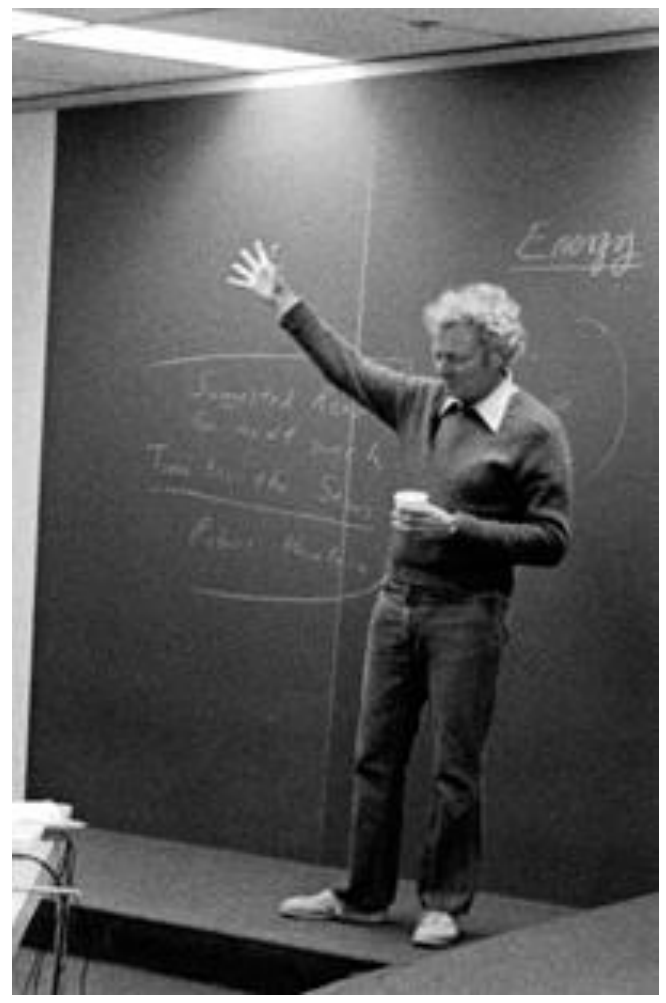

Figure 14: Leon Lederman talking to students.

## 7.2. CMS eLab

The CMS e-Lab allows students to analyze data to calibrate the detector and participate in discovery science. Calibrating the detector to "rediscover" previous measured results is an important part of the early scientific activity at CMS.

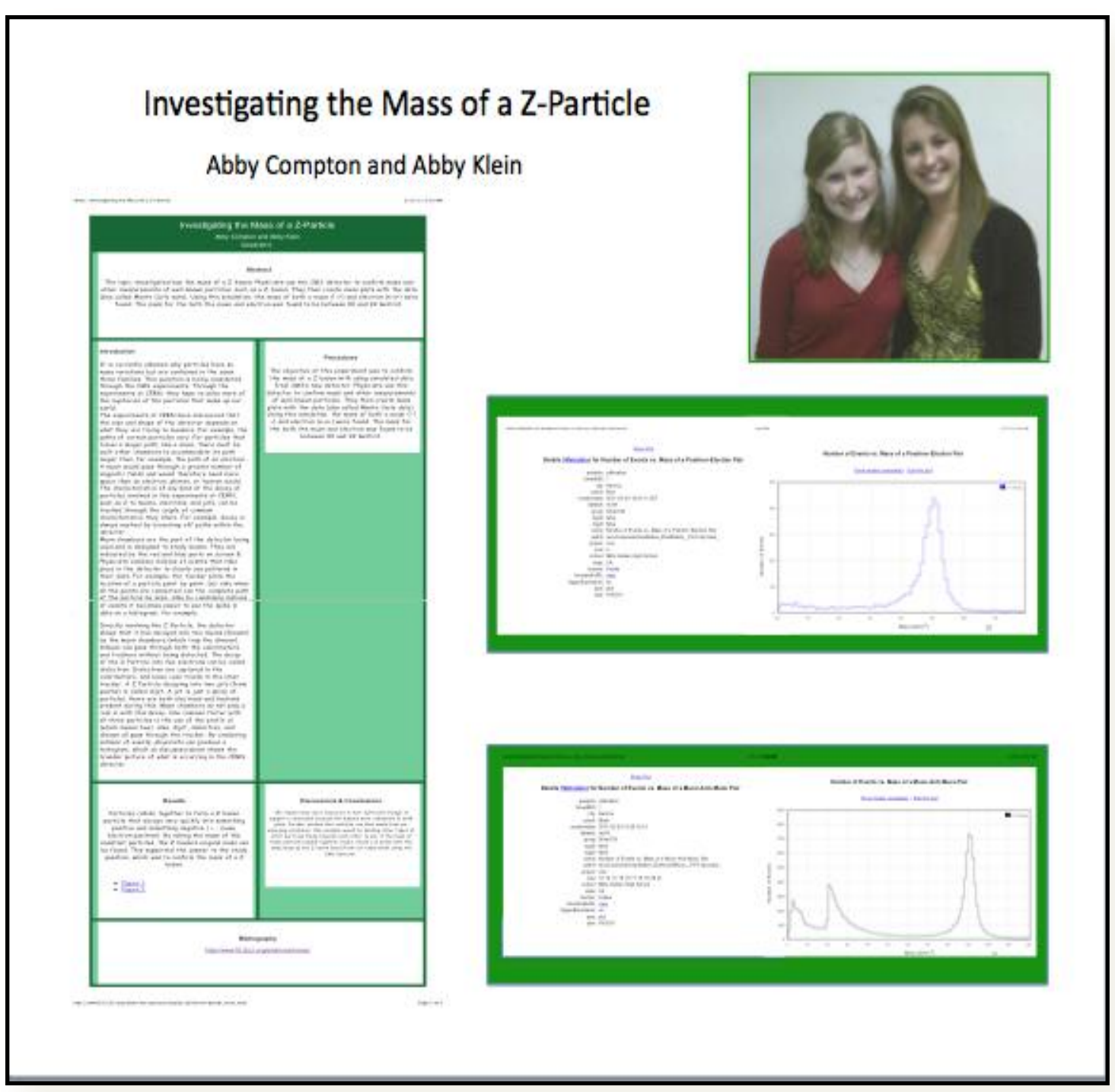

Figure 15: An example webpage of the CMS eLab

### 7.3. CMS Masterclass

Physicists look at CMS particle collisions in event displays to see what happened or in histograms of multiple events. On Masterclass Day, it is students who are physicists for a day. In 2011 the CMS Masterclass had 197 students in the U.S., 13 institutes in the U.S. plus approx 30 non-US institutes.

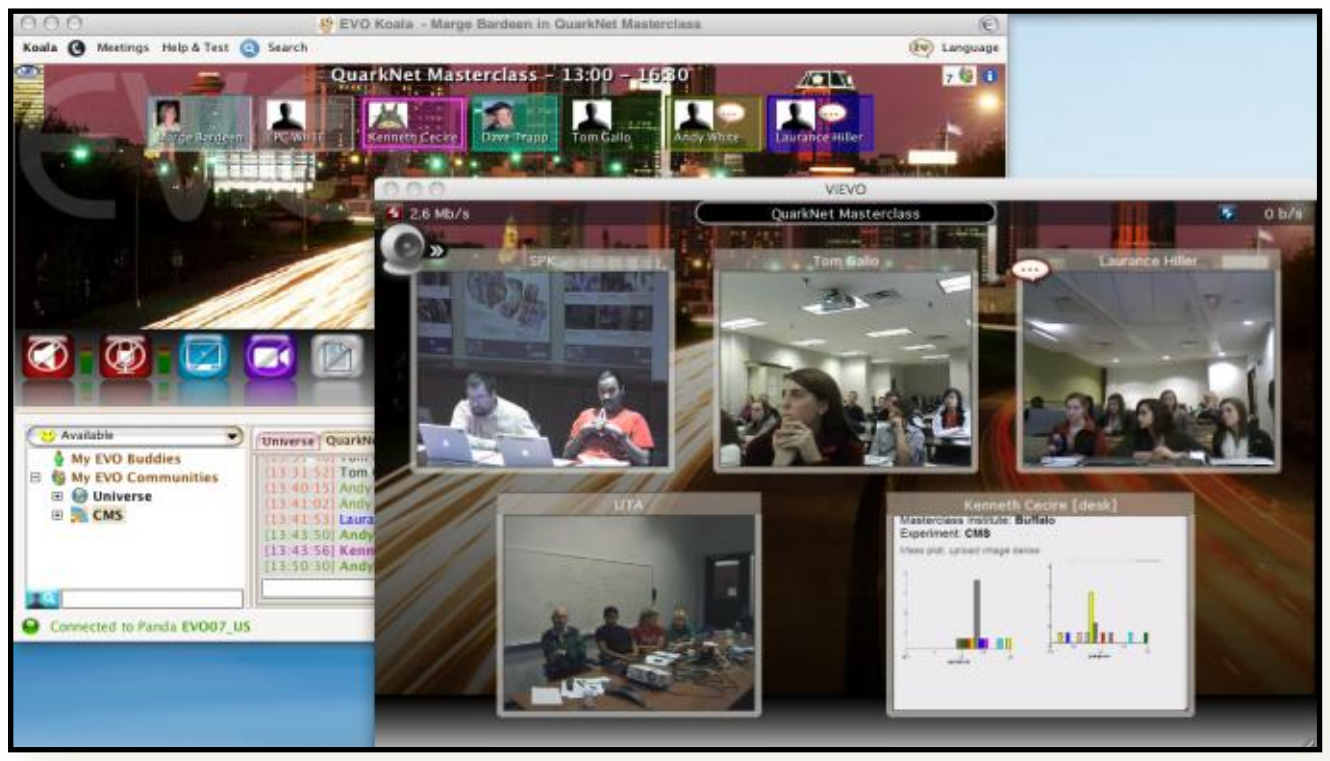

Figure 16: Monitoring the CMS Masterclass on EVO

### 7.4. IceCube Outreach Programs

The IceCube has an extensive and ongoing program of education and outreach. Some of the emphasis is on informal learning

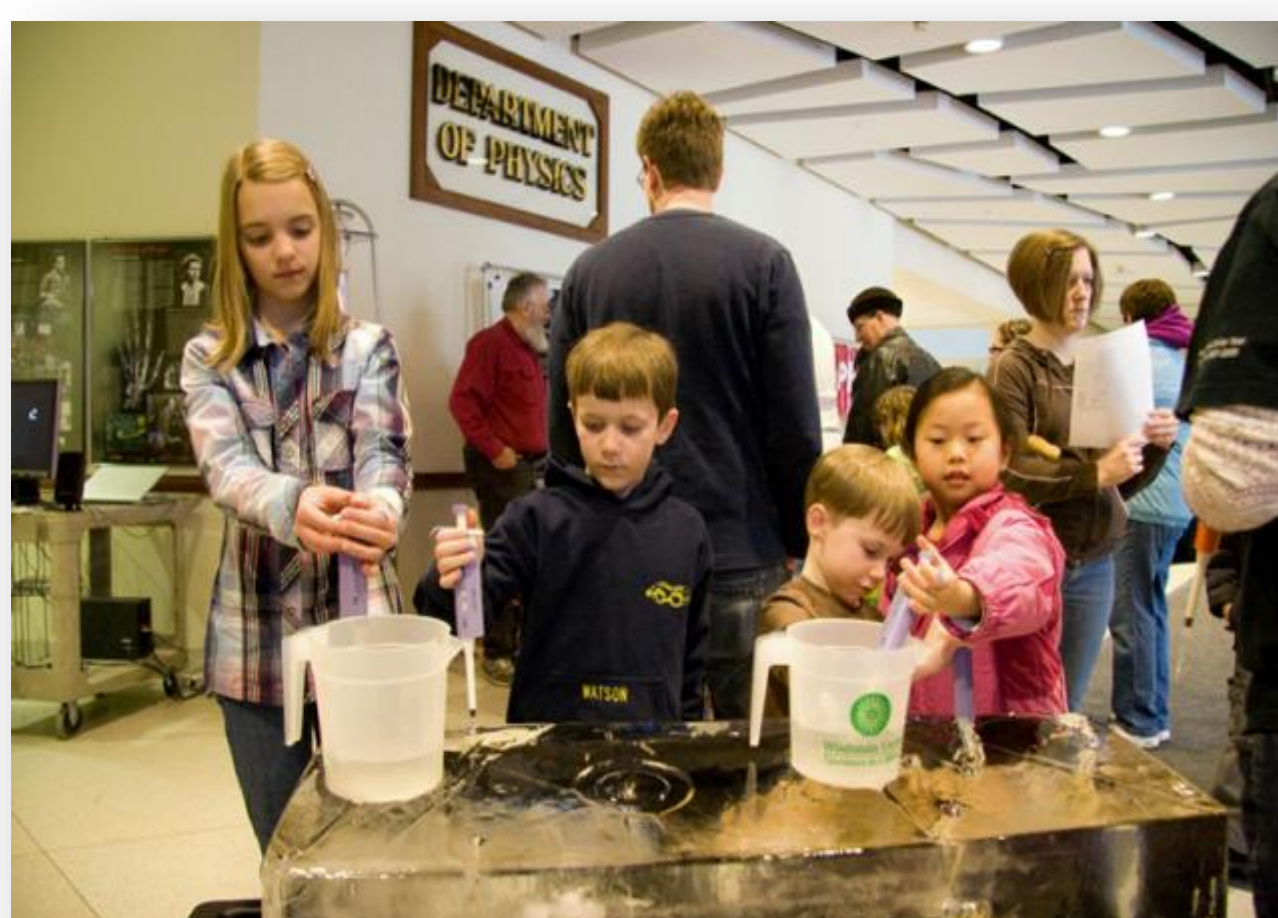

Figure 17: Informal learning by the IceCube collaboration

Some of the informal learning occurs at the Adler Planetarium in Chicago. Some 500,000 visitors per year see a dynamic display of real data. In Madison, Wisconsin, they participate in a hands-on ice drilling activity. It simulates the building of the IceCube detector and is appropriate for all ages

IceCube also has blogs and broadcasts from the South Pole, covering wintering over at the Pole, hot water drilling, and computing, running and maintaining the IceCube detector. Their curriculum covers: introduction to particle physics, access to real IceCube data, and generalization of concepts to illustrate abstract ideas.

### 7.5. Quarked!

Quarked! is a physics education project for ages 7 and up (based in at the University of Kansas). Project components include:

- Interactive web site (www.quarked.org) with animated videos, games, lesson-plans, FAQS, glossary, videos, etc.
- Hands on science programs for elementary and middle school students
- Links with QuarkNet, NSF informal science education and other grants

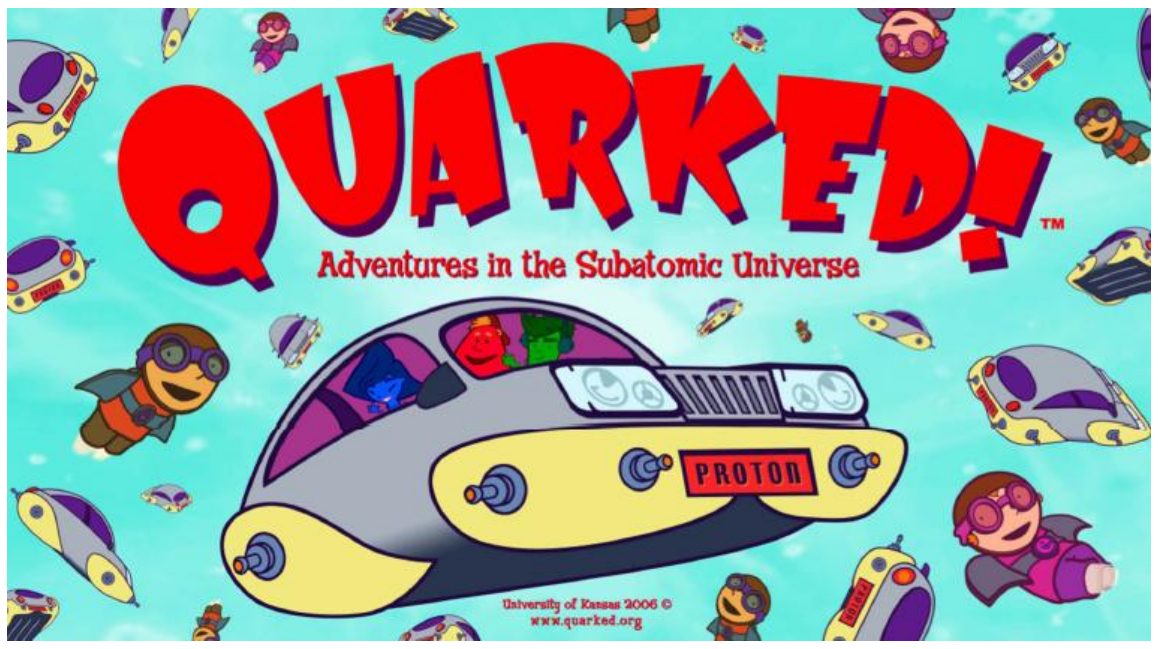

Figure 18: Homepage of Quarked!

The program's education programs are

- How Small is Small? (Grades 2 to 6)
- Quarks: Ups, Downs and the Universe (Grades 4 & up)

The unusual aspect of this program is the target audience. This age group was chosen because at "this stage young people are open to everything and don't know that physics is hard." The organizers believe that you can engage elementary & middle school aged children with concepts related to particle physics. They point out that over 5000 children participated from Kansas, Missouri, and Colorado in hands-on shows. Assessment has been done with students and their teachers and shows that the program is successful. The website continues to have over 5000 visitors per month.

### 7.6. Ligo Outreach Programs

The Ligo programs in a variety of outreach efforts and feature the Louisiana Science Education Center, which is a 5000 sq. foot Science Education Center. They conduct tours, present exhibits, and focus on K-12 and teacher development.

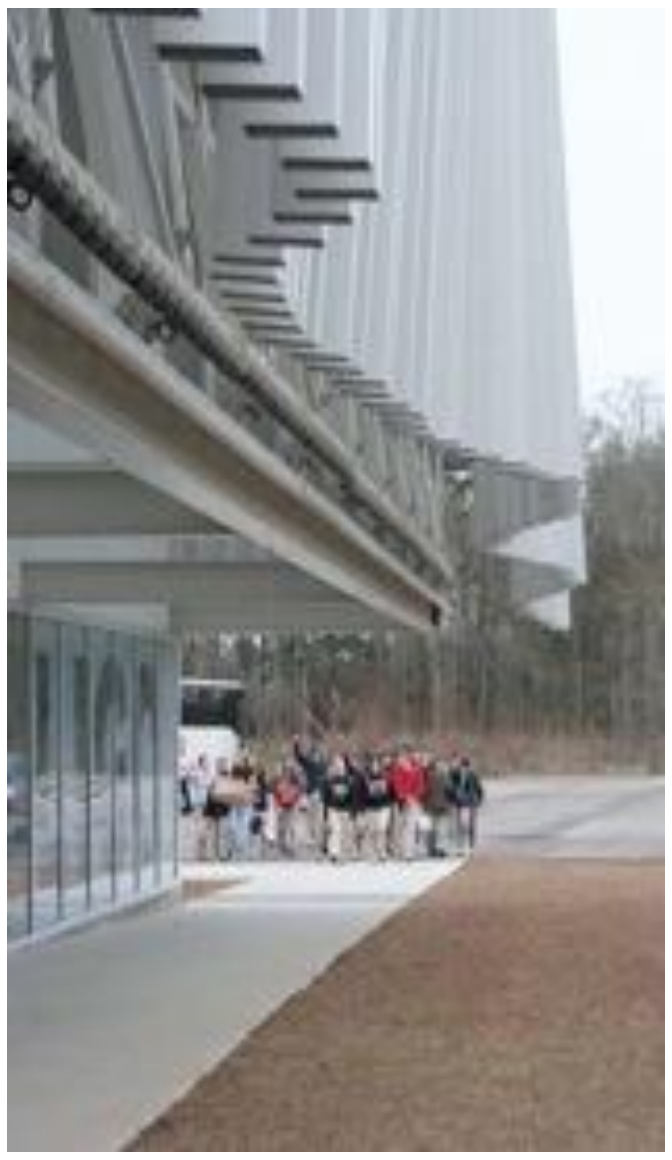

Figure 19: Ligo's Louisiana Science Education Center

### 7.7. Sanford Underground Lab Outreach Programs

The Sanford Underground Lab at the Homestake Mine in South Dakota aims at nationwide outreach. Their goals include exhibits, web features, games, blogs, student training, science fairs and festivals. Planning is underway for the Sanford Center for Science Education (SCSE):

– Leveraged by a substantial pledge from philanthropist T. Denny Sanford, the SCSE will be the education and outreach arm of DUSEL or its DOE equivalent
– The SCSE is envisioned to feature and leverage the science and engineering of the underground lab to excite and help prepare the next generation of scientists and engineers -- and also the broader citizenry.

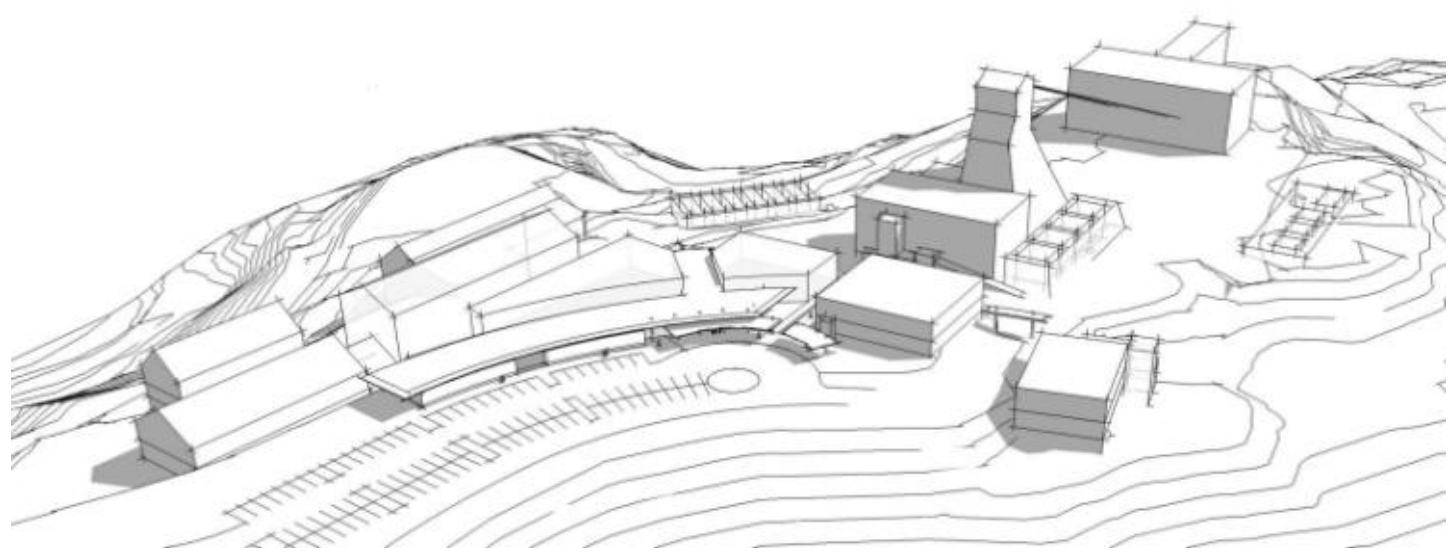

Figure 20: The campus of the Sanford Underground Lab

Early activities are building partnerships and capacity, and testing content for a variety of audiences. Special efforts are made to serve American Indian audiences, approximately 10% of the regional population.

### 7.8. MSU Planetarium Outreach Project

A planetarium outreach project at Michigan State University focuses on ATLAS at the planetarium. It is an outreach component of the NSF CAREER grant of Reinhard Schwienhorst. It aims to bring ATLAS, LHC, and HEP to planetarium audiences. Working in close cooperation with the Abrams Planetarium at MSU, they have already produced a 5-minute planetarium clip, which plays at the end of each public show. They are working on a full planetarium show to debut this fall.

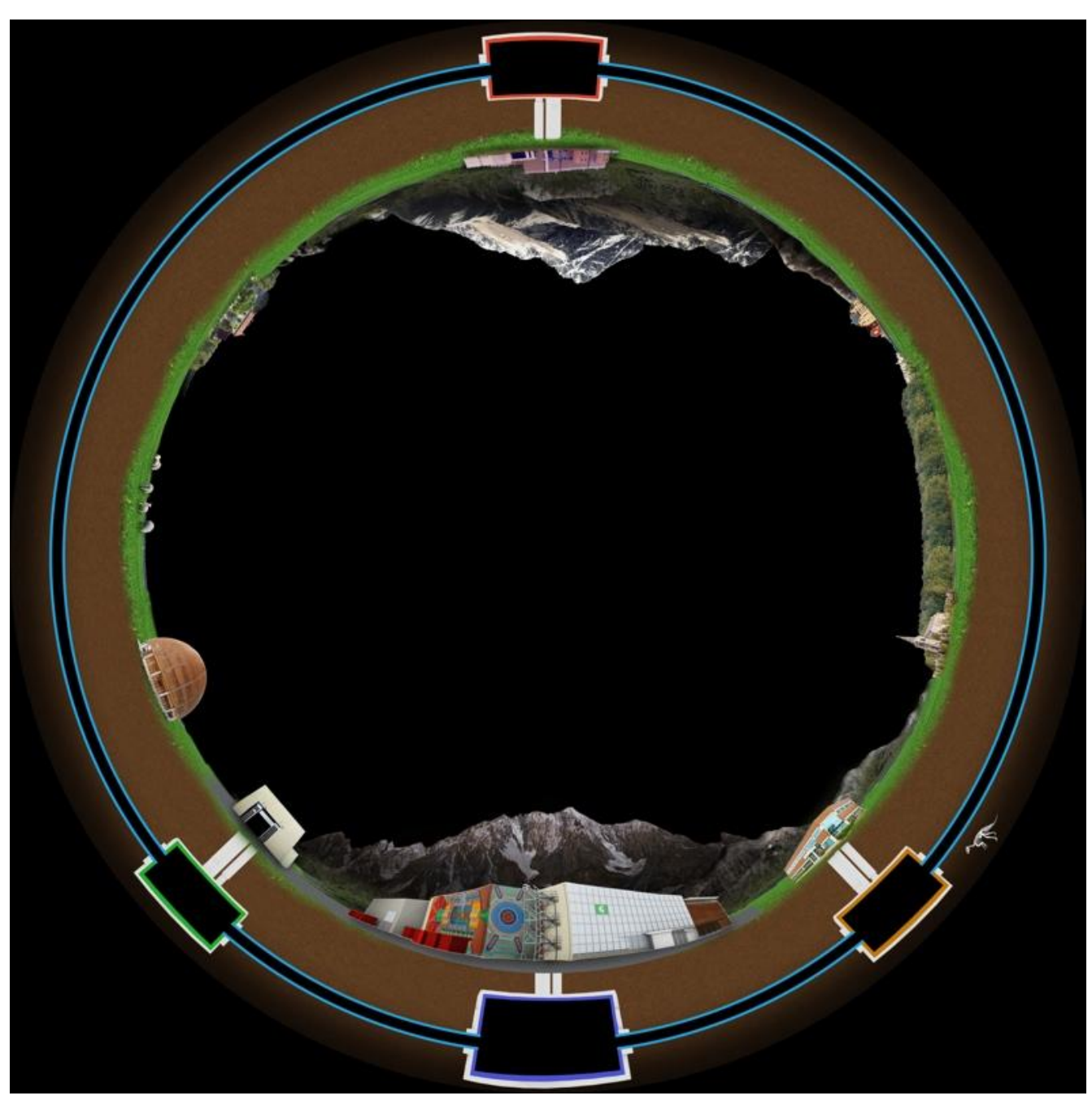

Figure 21: Example of a scene from the MSU planetarium show.

The planetarium display environment has both video projector and slide projectors. The show design is by MSU students: undergraduates, professors, graduate student,… with support from experts: a professional writing professor, a communication graphics and design professor, and a planetarium show developer

### 7.9. Pierre Auger Observatory Outreach Projects

The Pierre Auger Observatory is in Malargüe, Mendoza Province, Argentina. They have over 10 years of outreach in Mendoza and beyond, with a positive impact in all communities. The collaboration takes part in local traditions and fosters a sense of partnership in Auger's scientific mission. The Auger Visitor Center in its headquarters has hosted over 65,000 visitors since opening in 2001. The collaboration partnered with the Mendoza Province to construct the new James Cronin School in Malargüe in 2007

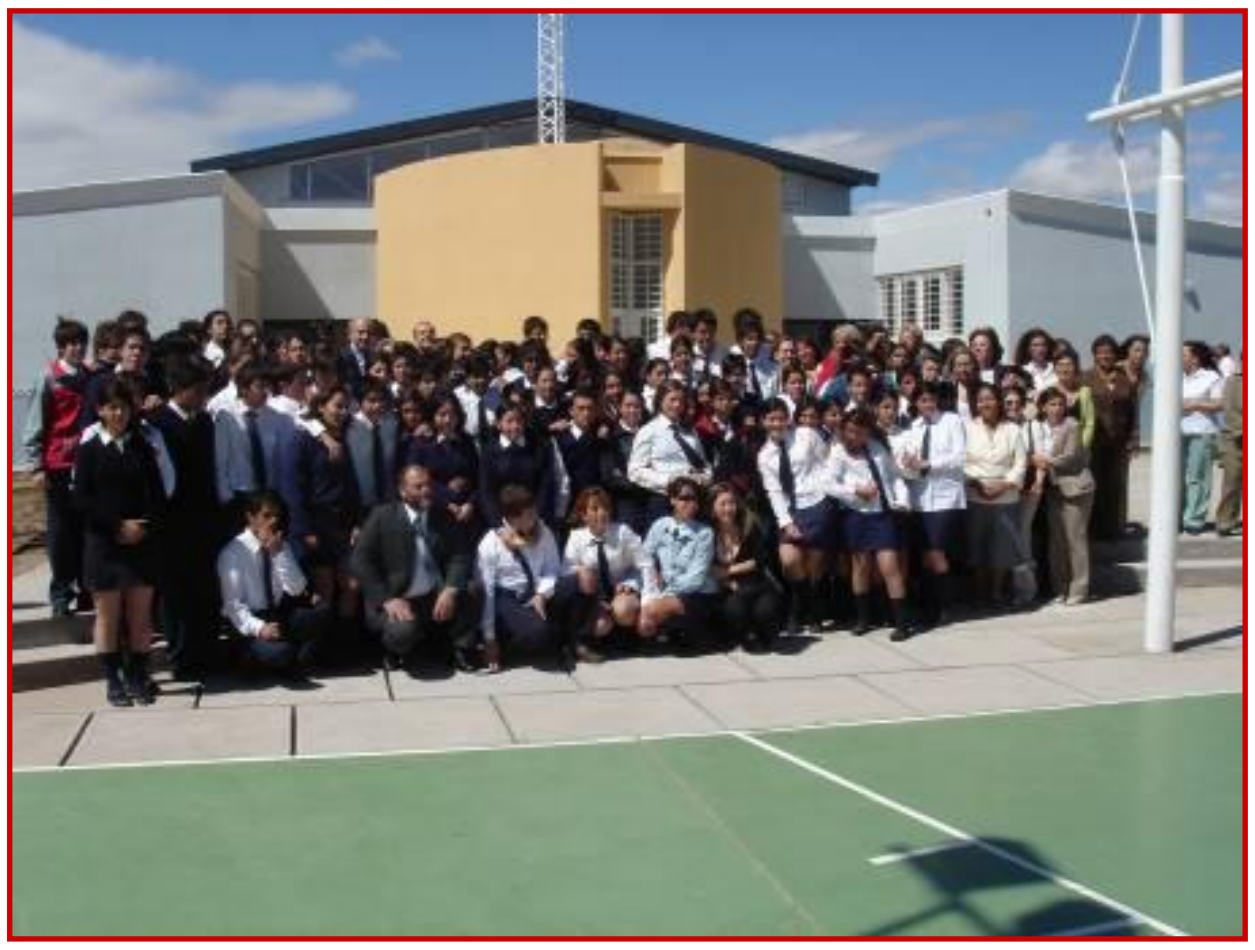

Figure 22: The new James Cronin School in Malargüe, Argentina

Their Rural Schools Program features observatory staff volunteer who bring cosmic ray science and infrastructure improvements to remote schools that cannot travel to Malargüe. The Collaboration sponsored 3-day Science Fairs at the Observatory in 2005, 2007, and 2010 drawing student and teacher teams from all over the Province. Collaborators serve as judges.

Students use reconstructed air shower data released online for science fairs and independent research projects. http://www.auger.org (5 languages, 4000 hits per year). Collaborators offer courses in hands-on science instruction to local teachers.

### 7.10. Plain English Summaries

The concept of descriptions of experimental results understandable by the public (or perhaps the scientifically literate public) dates back at least to some of the LEP experiments. The specific idea of "Plain English Summaries" of experimental papers seems to have originated with the DOE and CDF experiments at the Fermilab TeVatron Collider. The ATLAS Experiment at CERN has begun writing plain English summaries, including one asking: "Is nature supersymmetric?", one on dijet events, and one summarizing the results presented by ATLAS at the Lepton Photon Conference in Mumbai.

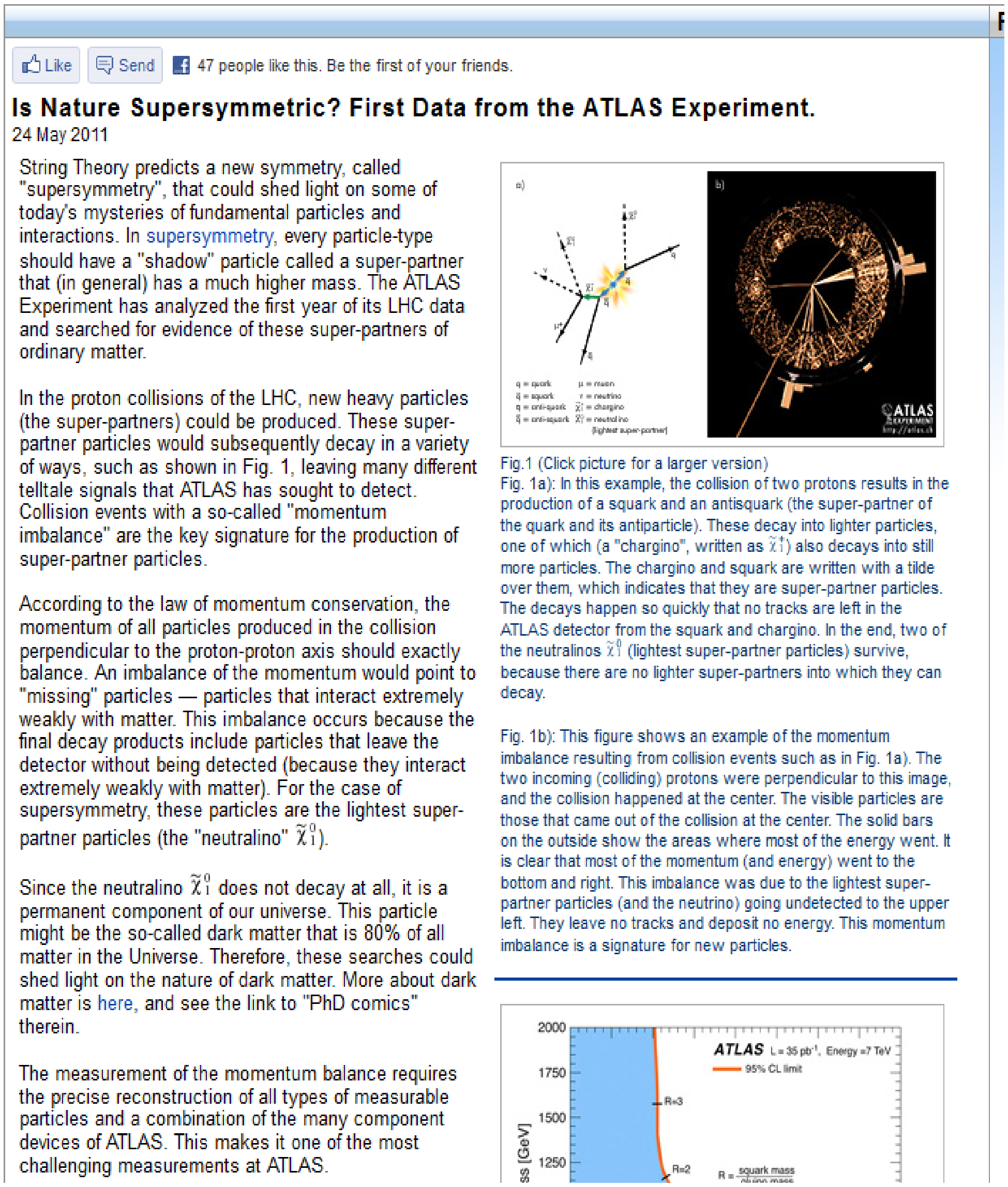

Like Send 47 people like this. Be the first of your friends.

**Is Nature Supersymmetric? First Data from the ATLAS Experiment.**

24 May 2011

String Theory predicts a new symmetry, called "supersymmetry", that could shed light on some of today's mysteries of fundamental particles and interactions. In supersymmetry, every particle-type should have a "shadow" particle called a super-partner that (in general) has a much higher mass. The ATLAS Experiment has analyzed the first year of its LHC data and searched for evidence of these super-partners of ordinary matter.

In the proton collisions of the LHC, new heavy particles (the super-partners) could be produced. These super-partner particles would subsequently decay in a variety of ways, such as shown in Fig. 1, leaving many different telltale signals that ATLAS has sought to detect. Collision events with a so-called "momentum imbalance" are the key signature for the production of super-partner particles.

According to the law of momentum conservation, the momentum of all particles produced in the collision perpendicular to the proton-proton axis should exactly balance. An imbalance of the momentum would point to "missing" particles — particles that interact extremely weakly with matter. This imbalance occurs because the final decay products include particles that leave the detector without being detected (because they interact extremely weakly with matter). For the case of supersymmetry, these particles are the lightest super-partner particles (the "neutralino" $\tilde{\chi}_1^0$).

Since the neutralino $\tilde{\chi}_1^0$ does not decay at all, it is a permanent component of our universe. This particle might be the so-called dark matter that is 80% of all matter in the Universe. Therefore, these searches could shed light on the nature of dark matter. More about dark matter is here, and see the link to "PhD comics" therein.

The measurement of the momentum balance requires the precise reconstruction of all types of measurable particles and a combination of the many component devices of ATLAS. This makes it one of the most challenging measurements at ATLAS.

Fig.1 (Click picture for a larger version)

Fig. 1a): In this example, the collision of two protons results in the production of a squark and an antisquark (the super-partner of the quark and its antiparticle). These decay into lighter particles, one of which (a "chargino", written as $\tilde{\chi}_1^\pm$) also decays into still more particles. The chargino and squark are written with a tilde over them, which indicates that they are super-partner particles. The decays happen so quickly that no tracks are left in the ATLAS detector from the squark and chargino. In the end, two of the neutralinos $\tilde{\chi}_1^0$ (lightest super-partner particles) survive, because there are no lighter super-partners into which they can decay.

Fig. 1b): This figure shows an example of the momentum imbalance resulting from collision events such as in Fig. 1a). The two incoming (colliding) protons were perpendicular to this image, and the collision happened at the center. The visible particles are those that came out of the collision at the center. The solid bars on the outside show the areas where most of the energy went. It is clear that most of the momentum (and energy) went to the bottom and right. This imbalance was due to the lightest super-partner particles (and the neutrino) going undetected to the upper left. They leave no tracks and deposit no energy. This momentum imbalance is a signature for new particles.

Figure 23: Example of a figure

## Acknowledgments

I would like to especially thank the speakers in the parallel session whose talks I have very quickly summarized (see their talks for all the details). This includes: Marge Bardeen, Alice Bean, Reinhard Schwienhorst, Peggy Norris, Greg Snow, Marco Cavaglia, James Madsen, Charles Timmermans, Helio Takai, Ken Cecire, Mike Fetsko, and Tom Jordan.

Supported by the Director, Office of Science, Office of High Energy and Nuclear Physics, the Division of High Energy Physics of the U.S. Department of Energy under Contract No. DE–AC02–05CH11231, and by the U.S. National Science Foundation under Agreement No. PHY-0652989;